# Trustworthy human-centric based Automated Decision-Making Systems

**Marcelino Cabrera[1], Carlos Cruz[2], Pavel Novoa-Hernández[2], David A. Pelta[2] and José Luis Verdegay[2,*]**

[1]Departamento de Lenguajes y Sistemas Informáticos, Universidad de Granada;
mcabrera@ugr.es
[2]Departamento de Ciencias de la Computación e I.A., Universidad de Granada;
{carloscruzcorona, pavelnovoa, dpelta}@ugr.es
*Correspondence: verdegay@ugr.es

**Abstract:** Automated Decision-Making Systems (ADS) have become pervasive across various fields, activities, and occupations, to enhance performance. However, this widespread adoption introduces potential risks, including the misuse of ADS. Such misuse may manifest when ADS is employed in situations where it is unnecessary or when essential requirements, conditions, and terms are overlooked, leading to unintended consequences. This research paper presents a thorough examination of the implications, distinctions, and ethical considerations associated with digitalization, digital transformation, and the utilization of ADS in contemporary society and future contexts. Emphasis is placed on the imperative need for regulation, transparency, and ethical conduct in the deployment of ADS.

## 1. Introduction

The widespread, rapid, and extensive development of digital service platforms, as well as debates about public data spaces and new technologies such as artificial intelligence (AI), affect all areas of our society. Many new ways of communicating, shopping, and accessing services or information online have been integrated into our daily lives and are constantly evolving. The internet does not play a central role in our lives; we simply cannot conceive of our lives without the internet.

But the internet and its future demand a key process: the digitalization of society, which also has several important implications as it enables faster and more efficient access to information, facilitates connectivity between people, devices, and systems, and fosters innovation by providing a platform for the development of new technologies and services, ranging from education to healthcare and manufacturing, improving not only efficiency but also creating new business models and opportunities.

As will become obvious, it is difficult to imagine a scenario in which the internet can function effectively without such a scenario being digitalized, and therefore digitalization and the digital transformation it entails is a central component for the development and continued evolution of the internet and its ability to positively impact society in a verifiable, secure, trustworthy, and human-centric way.

The European Digital Agenda [1] for the decade 2020-2030 addresses these issues by focusing on creating secure digital spaces and services, achieving a level playing field in digital markets with large platforms, and strengthening Europe's digital sovereignty while contributing to Europe's goal of climate neutrality by 2050. More specifically, it addresses the profound changes brought about by digital technologies, the essential role of digital services and markets, and the EU's new technological and geopolitical ambitions, pursuing the achievement of four objectives, among which, given the subject matter of this work, the most important is that by 2030 all essential public services should be available online.

If only for the sake of achieving this far-reaching target, the need to transform the typologies of public services is clear. As public services must be available online, it is imperative to make the appropriate cross-cutting digital transformations to achieve the new services and systems.

Digital transformation is the action of integrating (digital) information technologies in all areas of a system to increase the efficiency of all its internal processes and improve the services it offers, which implies a migration of methods and a change in the logic and mentality of the organization's operations [2].

If the system we want to transform already has the necessary digital equipment, there are two levels of action: digitalization and digital transformation itself. The first refers to creating or improving processes by taking advantage of digital technologies and previously digitized data. The second requires making organizational changes by exploiting the full potential of technology.

As is easy to understand, the speed at which work can be done at each of these levels of development is much slower than the speed at which technology advances. In this sense, the disruption that the arrival of AI is producing in all social spheres is particularly striking.

Since artificial intelligence (AI) is growing at a rate and intensity that are often unjustified and because of the (alleged) benefits that come with its application, everyone wants to showcase AI as a keystone of their brand identity, even though many times they don't know if it will be useful, if it will be the best solution, or if the person using it will be able to comprehend the answers it will provide to the problems it is meant to solve.

Thus, it is common to read that AI is being applied, or that a certain system works with AI when what is happening is that we are applying a technique that is used in AI. This is the case, for example, of neural networks, metaheuristics, deep learning, etc. The application of a technique commonly used in the field of AI to solve a problem does not imply in any case that the problem is solved with AI and much less, as is often read so many times, that AI is solving the problem at hand. The examples are many and well known and therefore we will not dwell on them here.

But in the same way that the word "artificial intelligence" is misused when it is not applied, it is misused when we employ standard software—which is devoid of AI altogether—to solve an issue, believing that just because they are using technology, they are using AI.

While the first misuse can be understood as a form of fraud, the second can be explained from the point of view of the ignorance of those who act in this way: without having a clear idea of what AI is and without distinguishing it from standard software, for various reasons that would be too long to enumerate, the user assumes that the system they are using works with AI, when in fact it does not.

Assuming there is no more than misinformation and no malicious intent, confusion is frequent when it is said that a Decision Support System (DSS) or an Expert System (ES) is used to make decisions (or to solve a problem), understanding that they are using AI techniques. Both tools are no more than standard computer systems, which in general do not incorporate the characteristics of an AI-based system, as is the case of Automated Decision-Making Systems (ADS).

Many times, the confusion comes from labeling these tools as "intelligent". However, as the concept of intelligence is not well defined [3], the debate around such labeling may be sterile. In turn, it leads us to stop and fix the technological concepts that we are going to use in the following.

Therefore, to minimize the risks that may be involved in the digital transformation of a certain system, be it judicial, health, or economic, it seems appropriate to understand the differences and similarities between DSS, ES, and ADS. For this reason, in what follows we begin by describing each of these systems, and then, in section 3, we analyze the phases that the digital transition of the system we are considering should consider and finish by exposing some undesirable consequences that the use of ADS can bring about and that should be considered to avoid possible problems.

## 2. Making decisions as human beings do

As stated in [4], AI studies computer programs (and in some cases also hardware equipment) designed by people who, given a complex objective, act in the physical or digital domain by perceiving the context in which they operate by acquiring data, interpreting that data, whether structured or not, reasoning about the knowledge or processing the information derived from that data, and deciding the best action or actions to be taken to achieve the objective pursued.

Therefore, an AI-enabled system is first and foremost rational. This rationality is achieved by perceiving the environment in which the system is immersed through some sensors, reasoning about what is perceived, deciding what is the best action to take, and acting accordingly through some actuators that can modify the context in which the system operates.

On the other hand, as a scientific discipline, AI includes several approaches and techniques such as Machine Learning (of which deep learning and reinforcement learning are its most relevant sub-disciplines), Computational Reasoning (which considers planning, scheduling, knowledge representation, and reasoning, search, and optimization) and robotics (comprising control, perception, sensors, and actuators, as well as the integration of all other techniques in cyber-physical systems).

However, since not all digital technologies use AI, artificial intelligence cannot exist without digital technology. For this reason, carrying out the digital transformation of a certain system does not mean evolving it towards the exclusive use of techniques, methodologies, or tools from the field of AI. If the transformation is done correctly, there will be areas that may include them, others that do not, and others in which, despite everything, the role that people have always played will remain unchanged, without any need or possibility of digitization.

Let us focus on a specific aspect: decision-making. It is evident that in this context, DSS, ES, and ADS can play a leading role. These tools can have as many versions and nuances as each area of the application permits, but they won't help achieve a proper digital transformation if each isn't used appropriately in each situation where it's called for, such as when an ADS is used to solve a problem when a spreadsheet would suffice.

That is why it is important to distinguish them, apply them correctly, and make it easier for the digital transformation to place them in those areas where they are appropriate, minimizing the risks that could arise from misuse or even improper use. To this end, it is necessary to know the basic characteristics of the most important tools for decision-making based on digital technologies. Let us analyze each of them in the following to place the subject in its proper coordinates.

A DSS is a software tool used to support the decision-making process of an individual or an organization. These systems usually include a set of techniques and methods to collect and analyze relevant information, present it clearly, and provide the decision-maker with recommendations or options for action based on this analysis.

More theoretically, taking into account that a structured problem has a context with elements and relationships between them that we can understand, in very general terms, a DSS can be defined as a system under the control of one or more decision-makers that assists them in their decision-making activity by providing them with an organized set of tools, aimed at structuring parts of the decision-making situation and improving the final effectiveness of the outcome of the decision they make [5].

The concept of DSS was born in the early 1970s and first appeared in papers by J.D. Little [6] and G.A. Gorry and M.S. Scott Morton [7]. Morton later coined the term DSS and developed a two-dimensional context in which computers would assist decision-making tasks in management problems. The term and the concept of DSS became established in the late 1970s [8].

DSSs are used in a wide variety of fields, such as medicine, engineering, finance, project management, and others. For example, a DSS could be used to evaluate different investment options and select the most suitable one for a company, or to help a physician choose the most appropriate treatment for a patient based on his or her clinical history and other relevant factors. DSSs can be designed to be used by humans or by other computers and can be programmed to make decisions on their own or to provide recommendations that are then evaluated and adopted by a person, the latter utility being the most common.

The constituents of a DSS may vary depending on the type of system and the purpose for which it has been designed. In general, however, the following common components can be mentioned:

- ***Database and model base***: It is essential to have a database that organizes the data in a logical hierarchy that represents the granularity of the data. It is also necessary to have a model base, which is the model version counterpart of the database. This model base stores and organizes the different models that the DSS will use in its analysis and is the component that differentiates a DSS from other information systems.
- ***Database, knowledge, and model management modules***: These components are responsible for analyzing the collected information and generating options or recommendations for action based on these analyses. While the first one works according to database standards, the second one applies the rules to deduce the information the user wants. The third allows the creation of decision models, provides a mechanism to connect the various models, and allows the management and manipulation of the model base.

- *Input and output interfaces*: The first is responsible for collecting and processing the information needed for the system and could include the user interface for entering data or connecting to external databases. The output interface serves to present the options or recommendations for action generated by the system in a clear and easy-to-understand way for the user.

In addition to these basic components, a DSS can also include other additional modules, such as a security module to protect information or a maintenance module to perform updates and improvements to the system.

An ES is a computer application that employs a set of rules based on expert knowledge to solve problems that require human expertise [5]. ESs mimic the process of reasoning based on the association of information, as people do to solve concrete problems in some very specific domain of knowledge. Thus, while a non-expert can use an ES to improve his or her ability to solve complex problems by simulating dialogues with experts in a particular field, experts can use an ES as if it were a highly knowledgeable assistant.

It is difficult to pinpoint the beginning of the development of ESs. However, mainly due to technological issues, it can be said that its official birth coincides in time with that of DSSs, in the early 1970s, evolving its definition, conception, and theoretical principles with its commercialization [9].

Like DSSs, ESs are used in a wide variety of areas, such as legal, medical, engineering, and economics, but not to the detriment of many others. Thus, for example, an ES in medicine could use the knowledge base of a particular physician to diagnose a disease or recommend a treatment. An ES in engineering could use the knowledge base of a particular engineer to design a structure, and similarly, an ES in the judicial context could be used to resolve small, neighborhood litigation. But in any case, the knowledge on which they rely in their decision-making is referenced to a particular experience and therefore they are not objective or neutral when deciding.

ESs can be very useful in situations where it is difficult to obtain expert advice in person, as they provide information and proposals based on expert knowledge and experience. However, ESs cannot fully replace human specialists and should be utilized cautiously because they rely on knowledge bases that are not required to be consensus in a particular field of application.

As with DSSs, the components of an ES may vary from application to application, but as a common denominator, they consist of the following modules:

- *User Interface*: This is the most crucial part of an ES. This component accepts the user's query in a readable form and passes it to the inference engine. It then displays the results to the user. In other words, it is an interface that helps the user communicate with the ES.
- *Inference engine*: It is the brain of the ES. The inference engine contains rules to solve a specific problem. It is related to the knowledge of the Knowledge Base. It selects facts and rules to apply when trying to answer a user query. It provides reasoning about the information in the Knowledge Base. It also helps to develop deductions on the considered problem to find the solution. This component is also useful in formulating conclusions.
- *Knowledge Base:* It is a repository of facts. It stores all the knowledge about the problem domain. It is like a large container of knowledge, which is obtained from different experts in a specific field.

As can be easily understood, the successful operation of an ES will depend mainly on the depth, accuracy, and precision of the knowledge it stores.

Since ES employs expert knowledge and experience to make choices and suggest courses of action, they are frequently seen as intelligent due to their constituent parts and roles. However, since they can't learn and adjust to novel or unexpected circumstances, their applicability is restricted, and they could act erratically in scenarios that aren't covered in the knowledge base.

Note that an ES is programmed with a specific knowledge base and can use this knowledge base to make decisions or proposals for action in situations that fit this knowledge base, i.e., limited to the expertise of the expert or set of experts that designed it. However, if a situation arises that is not included in the knowledge base of the system, the ES will not know how to react and will not be able to provide an adequate response.

Therefore, although ES can be useful in specific situations and can make decisions and proposals based on expert knowledge and experience, they are not as flexible or adaptive as other types of intelligent systems, such as machine learning systems or intelligent agents.

From a less practical perspective [10], it is important to keep in mind that, although thinking is generally not constrained, reasoning does necessitate certain rules to guarantee the validity of the conclusions that can be drawn. Such rules are usually expressed verbally, in the ordinary language that supports reasoning and is the basis on which experts act. However, reasoning is part of thinking, and its correctness is essential; people reason because we think, but computers, and therefore ES, reason without thinking, i.e., they are endowed with reasoning but not thinking, at least not to the same extent as people. Machines reason deductively using algorithms, while people also reason inductively and not only algorithmically.

Thus, to summarize we can say on the one hand that an ES can be seen as a specific type of DSS that is based on the knowledge and experience of an expert in a particular field. These systems are used to simulate the reasoning and decision-making of a human expert in a certain area of knowledge and are based on a set of rules and principles that have been programmed into the system by an expert in that field. On the other hand, a DSS can be any computer system that is used to support the decision-making process and does not necessarily have to rely on the knowledge and experience of an expert in a specific field. DSSs can be used to analyze and evaluate different options or alternatives based on available data and facts and can provide options or recommendations to a human user or another computer.

A more contemporary tool, closely tied to AI, is the Automated Decision Systems (ADS). These systems are primarily designed for automated decision-making, that is, using technological processes and without direct human involvement.

According to [11], automated decision-making is a (computational) process, including AI techniques and approaches, which, fed by inputs and data received or collected from the environment, can generate, given a set of predefined objectives, outputs in a wide variety of forms (contents, assessments, recommendations, decisions, predictions, etc.). This has two consequences. First, the concept of ADS includes algorithmic decision-making as well as AI-based decision-making. Second, automated decision-making can produce or deliver a myriad of outcomes and therefore, the resulting legal consequences and the legal regimes applicable to ADS differ and are potentially multiple and varied.

In the same line, authors in [12] define an ADS as any tool, software, system, process, process, function, program, method, model, and/or formula designed with or using computation to automate, analyze, assist, augment, and/or replace government decisions, judgments, and/or policy implementation. ADSs affect opportunities, access, freedoms, liberties, security, rights, entitlements, needs, behavior, residency, and/or status through prediction, scoring, analysis, classification, demarcation, recommendation, allocation, enumeration, classification, tracking, mapping, optimization, imputation, inference, labeling, identification, grouping, exclusion, simulation, modeling, evaluation, fusion, processing, aggregation and/or computation.

Therefore, as is evident, an ADS must necessarily incorporate four major components: the technology that enables its automated performance, a decision model base that incorporates as many theoretical aspects as necessary to be able to make decisions, a module with the algorithms and machine learning models that serve to characterize its behavior and a base of rules and facts that define the context to which the ADS is oriented in each case. In addition, these four main components must be coupled with the rule base that represents the knowledge of experts in the specific area in which the ADS is to be applied.

Table 1 summarizes the main features of ES, DSS, and ADS.

**Table 1.** Main features of Expert Systems, Decision Support Systems, and Automated Decision-Making Systems.

|  | **Expert Systems** | **Decision Support Systems** | **Automated Decision-Making Systems** |
|---|---|---|---|
| Purpose | Simulate the thinking process of human experts | Assist decision-makers in analyzing data and making informed choices | Make decisions without human intervention |
| Knowledge representation | Rule-based | Knowledge-based, data-driven, model-based | Machine learning-based |
| Application | Specific, well-defined problems in a particular domain | Complex, unstructured, or semi-structured problems | Real-time, large-scale problems |
| Decision-making process | Deductive reasoning | Analytical reasoning, simulation, optimization | Predictive analytics, machine learning algorithms |
| User interaction | Limited user interaction | Interactive user interface | Minimal user interaction |
| Learning ability | No learning ability | Limited learning ability | Self-learning from data |

## 3. Digitalization and digital transformation

As we mentioned in the introduction, when we decide to initiate the digital transformation of a system, we will have to move on two levels. The first will consider integrating the technology into the system that is being transformed, which will involve converting data into formats that are suitable for the newly implemented model. Thus, for example, if all our information is on a traditional medium, typically paper, it will have to be digitized. At the second level, the transformation of the system itself will have to be addressed, i.e., the organizational model, the form of communication with users, and the processes involved in the system transformation will have to be changed, taking advantage of the opportunities for improvement offered by digital technology.

Starting the entire digital transformation by deciding which digital technology tools we will use at the end for the normal operation of the system, is like starting a house from the roof. It does not mean that we do not think that the house should have a roof, but before we start to see the material that will have the roof we will put, we must have land for construction, permission to build, have a structure, etc.

Before reaching that point when it makes sense to start implementing advanced technological tools, whether AI-based or not, to provide the services that society demands, we need to distinguish between digitization and digital transformation.

In many areas, multiple projects have already been undertaken for the digitization of public administrations, ranging from the digitization of documents to the necessary training to take advantage of the applications developed and services offered, for the use of electronic mail, digital archiving, and many others that are increasingly in full development.

But the digital transformation of a system is more than just its digitization since it is a set of actions aimed at improving and modernizing the system's procedures and behavioral habits, which, by making use of digital technologies, seek to improve its strategic competitiveness in terms of modernization, efficiency, and effectiveness. The digital transformation of a system, whether legal, health, or educational, requires a review of its operating models, operations, and technological strategy, which entails a cultural change at all levels, not least of which is training since management and users must acquire solid training to operate with ease in the new transformed system and be able to take advantage of its full potential. Thus, digital transformation begins by fundamentally changing the organization and strategy of the corresponding system, with new operating models, new products and services, new channels of communication and contact with users, and, of course, new forms of relationship between its agents and managers at all levels.

Therefore, although digitization and digital transformation are strongly interrelated and can overlap, they are distinguishable from each other because the former concerns operations and processes, while the latter refers to the model and the interactions between the actors involved in each case, This requires a leader, a unified management that focuses on people, guarantees the interoperability of systems, multiple and secure access to databases, and avoids drifting into unforeseen or unknown scenarios due to contradictory decisions or decisions that are not well accepted by users and managers.

To put it briefly, the person in charge of the digital transformation must also be concerned with change management, which involves prioritizing people over technology. If a digital transformation does not have the agreement, flexibility, and complicity of the people to whom it is addressed and those who are responsible for its development, the probability that the process will fail is very high.

Likewise, this direction must follow a technology-independent strategy, which in any case must be adjusted to a governance scheme that allows important decisions on technologies to be taken at the appropriate level and in an integrated manner, taking special care to prioritize those projects based on technologies that add the greatest value to the system, since resources will generally be very limited, if not scarce. The monitoring of the development of the strategy, change management, the governance scheme, and the results provided by the established system of priorities must be measurable, to control the level of digital maturity that the system is reaching, taking care to analyze emerging technologies, such as Data Science, virtual assistants, augmented reality, learning analytics or AI itself.

In this new fully digitalized scenario, and because of the initially designed strategy, is where the necessary tools must be focused, on whether they are AI-based or not, the feasibility of their implementation, and how their continuous optimization, improvement actions, and adaptation to the current regulations that apply to them will be carried out.

In any problem-solving domain, for example in online dispute resolution, AI cannot be considered as if it were a magic bullet to solve any problem, whether it needs the help of AI and resorting to it in all areas, branches, and circumstances, even if the problem in question is well solved with other methods that do not require the use of AI-based systems. On the contrary, to develop an application based on AI techniques and methodologies, it is a priority to validate the convenience of using AI, and its viability and to define the field of application, with the practical recommendation that the more specialized it is, the better the results will be.

In any case, the implementation of the new digitally transformed system will require the use of digital tools according to the nature of the problem to be solved. In this sense, we can resort to standard software packages, e.g. conventional Excel sheets or similar, more sophisticated software tools, e.g. DSS or ES, or AI-based tools, i.e. ADS.

## 4. ADS in practice

When we address the issue of automated decision-making that impacts people, why do we use the term ADS and not directly use the term AI?

As we know, AI is a field of research and development in Computer Science that deals with machine learning and the automation of knowledge-related tasks. But that term suggests closeness to human intelligence and therefore evokes something akin to responsibility for the consequences of decisions made. For this reason, the term ADS, which we very often use instead, is usually better accepted as being more appropriate, giving it the value of a synonym.

It is often said that everything that is programmed is only as good as who made the program, and since ADSs are built by people, these systems will be subject to the same feelings and constraints that their designers are subject to in their daily lives, albeit unconsciously.

From that point of view, the use of ADS can lead to unfair, even discriminatory, outcomes that can have negative effects on people's lives. ADS are neither neutral nor objective. They must be understood as systems that arise in a particular social context and are influenced by the structures and relationships (not excluding those of power) that prevail there.

The unconscious biases, structural inequality, and discrimination that are common in societies also impact ADS because these systems reflect the assumptions, values, perspectives, prejudices, and biases of the people who develop and apply them. This is because people make decisions at different times in the development and application of these systems, as well as in contexts different from those that might have been initially envisioned by their designers [13]. In the worst case, these decisions lead to unwanted discrimination.

So, it is evident that issues with understanding and accepting the judgments produced by the system might occasionally occur when ADSs, as well as the algorithms and programs on which they are built, are created using AI approaches and methodologies. We may feel threatened because we believe that these machines do our jobs better than us, or because no one knows how to explain their behavior when they act outside of what is expected, producing undesirable biases, unsustainable solutions, or, in short. After all, we consider that they do not behave ethically.

But ADSs use algorithms, which are nothing more than a specific form of instructions leading to the solution of a mathematical problem. Algorithms describe a solution path, perfectly interpretable by the computer, i.e. free of ambiguities, which computes the correct solution in finite time for each possible input defined for the given problem. In short, ADSs make decisions based on models, always acting based on algorithms.

Thus, ADSs can support or propose decisions by making recommendations or processing data. For example, an ADS could be used to build a list of candidates that meet certain criteria from many applicants for an apartment. From this list, a person must decide whom to invite to an apartment inspection. If the execution of a decision is delegated entirely to an ADS, no human person will be involved in the formulation of the outcome, and problems may arise from disagreements with the proposed solution.

The operation of ADSs is influenced by two types of algorithms that should be distinguished: rule-based algorithms and learning algorithms. In a rule-based algorithm, a programmer explicitly programs the instructions that execute the algorithm. On the other hand, learning algorithms attempt to "learn" rules from pre-existing data that lead to a desired outcome in an initial phase. People can help in this phase by providing the available data for training, for example by marking the data containing the desired result. In the next phase, the algorithm "applies" the learned rules to similar but unknown data to obtain a result. In this case, the data that have been used in the first phase have a great influence on the learned rules and thus on the result. But as the selection made in the first phase is subjective, the results achieved will obey that previous subjectivity.

Regardless of the algorithms used, and as previously referred to [13], the context must be considered to better understand ADSs, since it makes perfect sense to consider ADSs as systems belonging to a specific ecosystem, thus incorporating social components in addition to technical ones in their analysis. This specific historical, economic, and social context influences ADSs and characterizes their performance. But the systems, in turn, influence specific groups and individuals in that society, in that ecosystem, as well as the development of society. This influence can be negative and sometimes lead to disadvantages or discrimination, as described in [14].

However, that is the version that is seen from the side of the AI user, i.e. who receives the result of the application of that ADS. Because from the side of the designer of that system, sometimes the results conform exactly to what was consciously foreseen in the source algorithm and the corresponding software that reaches the users. The range of possibilities is more than wide, going from misbehavior due to fortuitous errors, undesirable but unavoidable, and which in any case must be reckoned with, to the self-interested marketing of ADSs in which their proper use is not assured in any way or, even worse, are not even based on AI (there is an increasing proliferation of products with the label "based on AI" that nobody knows what it consists of, nor what it is, nor how it works, but which acts as an advertising claim).

In any case, and from a very general point of view, it is obvious that the ethical level of certain AI-based software is context-dependent, meaning that while in some scenarios an action may be considered unethical, in others it may be considered suitable, giving rise to contradictory situations that are difficult to compare because we do not have tools that allow us to assess the ethical behavior of an AI-based system, nor, what is worse, sufficient legislation to regulate it.

Nevertheless, regardless of the possible implementation of the future European law on IA, progress is beginning to be made along these lines. Taking Spain as an example, a comprehensive law for equal treatment and non-discrimination has recently been enacted, which in article 23, on AI and automated decision-making mechanisms, states [15]:

Within the framework of the National AI Strategy, the Digital Bill of Rights, and European initiatives on AI, public administrations will favor the implementation of mechanisms so that the algorithms involved in decision-making used in public administrations consider criteria of minimization of bias, transparency, and accountability, whenever technically feasible. These mechanisms will include their design and training data and will address their potential discriminatory impact. To this end, impact evaluations to determine potential discriminatory bias will be promoted.

Public administrations, within the framework of their competencies in the field of algorithms involved in decision-making processes, will prioritize transparency in the design and implementation and the interpretability of the decisions adopted by them.

Public administrations and companies shall promote the use of ethical and reliable AI that respects fundamental rights, especially following the recommendations of the European Union in this regard.

An algorithm quality seal will be promoted.

Concern for all the versions that can be taken by the multiple and different forms of biases, discriminations, dysfunctions, etc. is not only legitimate but an obligation that concerns us all. That is why it is an absolute priority that, together with the digital transformation processes that are being carried out, when these transformations involve the use, in any sense, of AI-based systems, typically ADS, in all cases they are regulated and controlled by Supervisory Agencies that can act with fully organic and functional independence from the Public Administrations, in an objective, transparent and impartial manner, carrying out measures aimed at minimizing significant risks to people's health and safety, as well as to their fundamental rights, that may arise from the use of ADS, and that are responsible for the development, supervision, and monitoring of projects framed within the strategies for the advancement of AI that each government has established as well as, in the European case, those promoted by the EU, in particular those related to the regulatory development of AI and its possible uses in general.

## 5. Conclusions and lines of reflection

In short, and by way of the conclusion of the previous reflections, it is worth highlighting the following:

Social leaders must enhance the training and empowerment of citizens in the use of information and communication technologies especially in the use of AI. Although the advantages that digital transformation brings to social development and the evolution of public administrations are no longer in dispute, the ease with which people can take advantage of the benefits of administrative digitalization offers many doubts, since in many cases users lack a minimum training to properly use the corresponding tools, favoring the creation of growing social gaps due to lack of digital training.

In the same way that administrative transparency is demanded at all levels of public management, calling for ethical behavior, efforts must be redoubled to ensure that this is also true in the case of ADSs, among other reasons to try to minimize the risks of harm to individuals that their unconscious, indiscriminate or biased use may entail. Along these lines, promoting the use of free data and software in the design, implementation, and operation of ADSs should be mandatory.

The arrival of AI in society has been very rapid, having to face situations not previously imagined. But the truth is that AI has arrived and will stay with us forever. We must therefore learn to live with it, even if unforeseen disruptions may arise due to the lack of an appropriate legal framework. However, regulating a world that does not exist is very difficult. An ADS must have the capacity to learn and improve over time, and this makes it very difficult to regulate. But this regulation is as essential as its continuous and rapid updating, which is why there is a need for the creation of ADS Supervisory Agencies, made up of interdisciplinary teams of experts to deal with all these regulatory aspects and their dynamic updating.

The digitalization, transformation, and exploitation using ADS of any administrative, productive, or social framework must completely guarantee the security and privacy of user data. Interesting examples of the importance of the correct use of personal data can be found in the educational, military, sports, health, or judicial fields, among many others, highlighting the importance that these aspects have and will have in the future for the assurance of well-being and personal privacy.

## Acknowledgement

This research has been partially funded by projects PID2020-112754GBI0, MCIN/AEI /10.13039/501100011033, B-TIC-640-UGR20, FEDER/ Regional Government of Andalusia-Spanish Ministry of Economic Transformation, Industry, Knowledge, and Universities and TED2021-131127B-I00, Spanish Ministry of Science and Innovation.

## Conflicts of Interest:

The authors declare no conflict of interest.